\documentclass[runningheads]{llncs}
\usepackage{graphicx}
\usepackage{hyperref}
\usepackage{color}
\usepackage{multirow}

\begin{document}
\title{Brain tumor segmentation with self-ensembled, deeply-supervised 3D U-net neural networks: a BraTS 2020 challenge solution.}
\titlerunning{BraTS 2020: self-ensembled, deeply-supervised 3D U-Net CNNs}
%
\author{Théophraste Henry \inst{1} * \and Alexandre Carré \inst{1} * \and Marvin Lerousseau \inst{1,2} \and Théo Estienne\inst{1,2} \and Charlotte Robert\inst{1,3} \and Nikos Paragios\inst{4} \and Eric Deutsch\inst{1,3}}
\authorrunning{T. Henry et al.}
%
\institute{Université Paris-Saclay, Institut Gustave Roussy, Inserm, Radiothérapie Moléculaire et Innovation Thérapeutique, F-94805, Villejuif, France.  \and Université Paris-Saclay, CentraleSuplec, 91190, Gif-sur-Yvette, France \and Gustave Roussy, Département d'oncologie-radiothérapie, F-94805, Villejuif, France \and Therapanacea, Paris, France\\}
%
\maketitle
\let\thefootnote\relax\footnotetext{$^*$ equally contributing authors}
\begin{abstract}
Brain tumor segmentation is a critical task for patient’s disease management. In order to automate and standardize this task, we trained multiple U-net like neural networks, mainly with deep supervision and stochastic weight averaging, on the Multimodal Brain Tumor Segmentation Challenge (BraTS) 2020 training dataset. Two independent ensembles of models from two different training pipelines were trained, and each produced a brain tumor segmentation map. These two labelmaps per patient were then merged, taking into account the performance of each ensemble for specific tumor subregions. Our performance on the online validation dataset with test time augmentation were as follows: Dice of 0.81, 0.91 and 0.85; Hausdorff (95\%) of 20.6, 4,3, 5.7 mm for the enhancing tumor, whole tumor and tumor core, respectively. Similarly, our solution achieved a Dice of 0.79, 0.89 and 0.84, as well as Hausdorff (95\%) of 20.4, 6.7 and 19.5mm on the final test dataset, ranking us among the top ten teams. More complicated training schemes and neural network architectures were investigated without significant performance gain at the cost of greatly increased training time. Overall, our approach yielded good and balanced performance for each tumor subregion. Our solution is open sourced at 
\url{https://github.com/lescientifik/open_brats2020}


\end{abstract}

\section{Introduction}
\subsection{Clinical overview}
Gliomas are the most frequent primitive brain tumors in adult patients and exhibit various degrees of aggressiveness and prognosis. Magnetic Resonance Imaging (MRI) is required to fully assess tumor heterogeneity, and the following sequences are conventionally used: T1 weighted sequence (T1), T1-weighted contrast enhanced sequence using gadolinium contrast agents (T1Gd), T2 weighted sequence (T2), and fluid attenuated inversion recovery (FLAIR) sequence. 

\begin{figure}
    \centering
    \includegraphics[width=8cm]{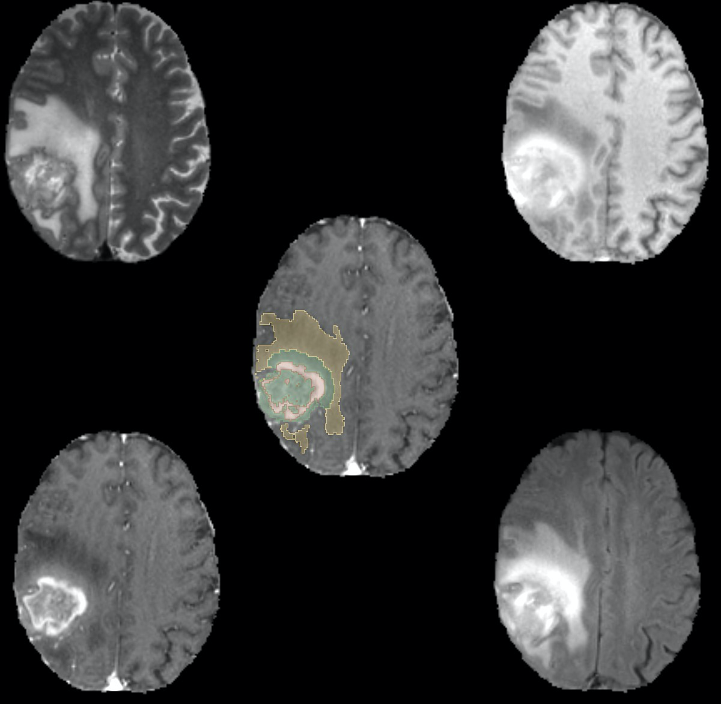}
    \caption{Example of a brain tumor from the BraTS 2020 training dataset. \textbf{Red}: enhancing tumor (ET), \textbf{Green}: non enhancing tumor/ necrotic tumor (NET/NCR), \textbf{Yellow}: peritumoral edema (ED). Upper Left: T2 weighted sequence, Upper Right: T1 weighted sequence, Lower Left: T1-weighted contrast enhanced sequence, Lower Right: FLAIR sequence Middle: T1-weighted contrast enhanced sequence with labelmap overlay} 
    \label{fig1}
\end{figure}

Four distinct tumoral subregions can be defined from MRI: the “enhancing tumor” (ET) which corresponds to area of relative hyperintensity in the T1Gd with respect to the T1 sequence; the “non enhancing tumor” (NET) and the “necrotic tumor” (NCR) which are both hypo-intense in T1-Gd when compared to T1; and finally the “peritumoral edema” (ED) which is hyper-intense in FLAIR sequence. These almost homogeneous subregions can be clustered together to compose three “semantically” meaningful tumor subparts: ET is the first cluster, addition of ET, NET and NCR represents the “tumor core” (TC) region, and addition of ED to TC represents the “whole tumor” (WT). Example of each sequence and tumor subvolumes is provided in Figure \ref{fig1} using 3D Slicer \cite{fedorov_3d_2012}.

Accurate delineation of each tumor subregion is critical to patient’s disease management, especially in a post-surgical context. Indeed, the radiation oncologist is required to segment the tumor, including the surgical resection cavity, the residual enhancing tumor and surrounding edema according to the Radiation Therapy Oncology Group (RTOG) \cite{niyazi2016estro}. Correct segmentation could also unveil prognostic factors through the use of radiomics or deep-learning based approach \cite{dercle_reinventing_2020}.

\subsection{Multimodal Brain Tumor Segmentation challenge 2020}

The Multimodal Brain Tumor Segmentation Challenge 2020 \cite{menze_multimodal_2015,bakas_advancing_2017,bakas_identifying_2019,bakas_s_segmentation_2017,bakas_s_segmentation_2017-1} was split in three different tasks: segmentation of the different tumor sub-regions, prediction of patient overall survival (OS) from pre-operative MRI scans, and evaluation of uncertainty measures in segmentation.
The Segmentation challenge consisted in accurately delineating the ET, TC and WT part of the tumor. The main evaluation metrics were an overlap measure and a distance metric. The commonly used Dice Similarity Coefficient (DSC) measures the overlap between two sets. In the context of ground truth comparison, it can be defined as follows:

\begin{equation}
DSC = \frac{2TP}{2TP+FP+FN}
\end{equation}

\noindent with TP the true positives (number of correctly classified voxels), FP the false positives and FN the false negatives. It is interesting to note that this metric is insensitive to the extent of the background in the image.
The Hausdorff distance \cite{huttenlocher_comparing_1993} is complementary to the Dice metric, as it measures the maximal distance between the margin of the two contours. It greatly penalizes outliers: a prediction could exhibits almost voxel-perfect overlap, but if a single voxel is far away from the reference segmentation, the Hausdorff distance will be high. As such, this metric can seem noisier than the Dice index, but is very handy to evaluate the clinical relevance of a segmentation. As an example, if a tumor segmentation encompasses distant healthy brain tissue, it would require manual correction from the radiation oncologist to prevent disastrous consequences for the patient, even if the overall overlap as measured by the Dice metric is good enough.

\section{Methods}

Two independent training pipelines were designed, with a common neural network architecture based on the 3D U-Net with minor variations (described below). These two different training approaches were kept separate in order to promote network predictions’ diversity. The specific details of each pipeline will be described below, and referred to as pipeline A and pipeline B. 

\subsection{Neural network architecture}

\begin{figure}
 \centering
 \includegraphics[width=8cm]{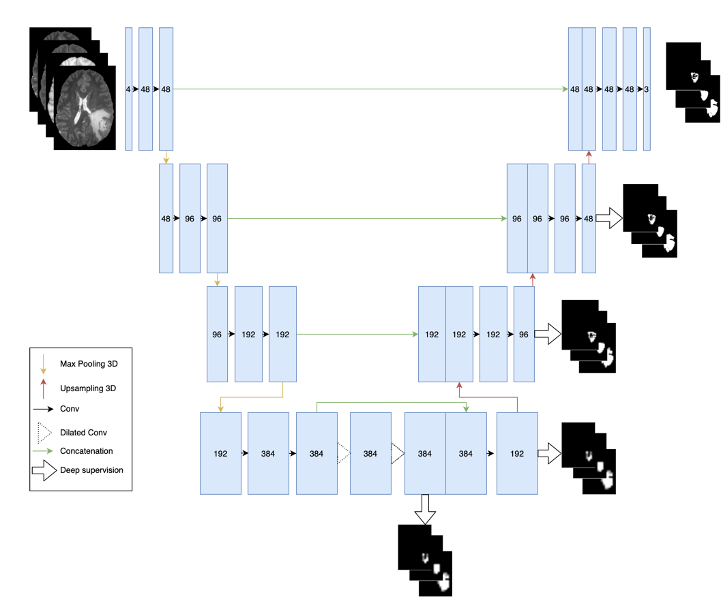}
 \caption{Neural Network Architecture: 3D U-Net \cite{cicek_3d_2016} with minor modifications} 
 \label{fig2}
\end{figure}

After neural network architecture exploration, the chosen network used an encoder decoder architecture, heavily inspired by the 3D U-Net architecture from Çiçek et al \cite{cicek_3d_2016}. The architecture used is displayed in Figure \ref{fig2}.

In the following description, a stage is defined as an arbitrary number of convolutions that does not change the spatial dimensions of the feature maps. All convolutions were followed by a normalization layer and a nonlinear activation (ReLU layer \cite{nair_rectified_nodate}). Group normalization \cite{wu_group_2018} (A) and Instance normalization \cite{ulyanov_instance_2017} (B) were used as a replacement for Batch Normalization \cite{ioffe_batch_2015} due to a small batch size during training and good theoretical performance on non-medical datasets.

The encoder had four stages. Each stage consisted of two 3x3x3 convolutions. The first convolution increased the number of filters to the predefined value for the stage (48 for stage 1), while the second one kept the number of output channels unchanged.  Between each stage, spatial downsampling was performed by a MaxPool layer with a kernel size of 2x2x2 with stride 2. After each spatial downsampling, the number of filters was doubled. After the last stage, two 3x3x3 dilated convolutions with a dilation rate of 2 were performed, and then concatenated with the last stage output.

The decoder part of the network was almost symmetrical to the encoder. Between each stage, spatial upsampling was performed using a trilinear interpolation. Shortcut connections between encoder and decoder stages that shared the same spatial sizes were performed by concatenation. The decoder stage performing at the lowest spatial resolution was made up of only one 3x3x3 convolution. Last convolutional layer used a 1x1x1 kernel with 3 output channels and a sigmoid activation.

The previous winner of the Brats challenge \cite{noauthor_two-stage_nodate} limited their downsampling steps to 3. We hypothesized that further downsampling of the features maps, given the limited size of the input (128x128x128), would lead to irreversible loss of spatial information. As the last stage of the encoder takes much less GPU memory than the first, the dilation trick \cite{chen_deeplab_2017} was used to perform a pseudo fifth stage at the same spatial resolution as the fourth stage.

3D attention U-Nets were also trained, using the Convolutional Block Attention Module \cite{woo_cbam_2018} added at the end of each encoder stage.

\subsection{Loss Function}

Inspired by the conciseness of the 2019 winning solution \cite{noauthor_two-stage_nodate}, the neural network was trained using only the Dice Loss \cite{milletari_v-net_2016} (A). The loss L is computed batch-wise and channel-wise, without weighting:

\begin{equation}
DSC = 1 - \frac{1}{N}\sum_{n}\frac{S_{n}*R_{n} + \varepsilon}{S_{n}^{2} + R_{n}^{2} + \varepsilon}
\end{equation}

\noindent with n the number of output channels, S the output of the neural network after sigmoid activation, R the ground truth label and $\epsilon$ a smoothing factor (set to 1 in our experiment). For diversity, the pipeline B used a slightly different formulation of the Dice Loss, without squaring the terms of the denominator. Similarly, optimization was made directly on the final tumor regions to predict (ET, TC and WT) and not on their components (ET, NET-NCR, ED). The neural network output was a 3-channel volume, each channel representing the probability map for each tumor region.

Deep supervision \cite{wang_training_2015} was performed after the dilated convolutions, and after each stage of the decoder (except the last) as in \cite{qin_basnet_2019}. Deep supervision was achieved by adding an extra 1x1x1 convolution with sigmoid activation and trilinear upsampling. Like the main output, each of this additional convolution resulted in a 3-channel volume, each channel representing the probability map for each tumor region (ET, TC and WT). The final loss was the unweighted sum of the main output loss, and the four auxiliary losses.

\subsection{Image pre-processing}

Since MRI intensities vary depending on manufacturers, acquisition parameters, and sequences, input images needed to be standardized. Min-max scaling of each MRI sequence was performed separately, after clipping all intensity values to the 1 and 99 percentiles of the non-zero voxels distribution of the volume (A). Pipeline B performed a z-score normalization of the non-zero voxels of each IRM sequence independently. 

Images were then cropped to a variable size using the smallest bounding box containing the whole brain, and randomly re-cropped to a fixed patch size of 128x128x128. This allowed to remove most of the useless background that was present in the original volume, and to learn from an almost complete view of each brain tumor.

\subsection{Data augmentation techniques}

To prevent overfitting, on-the-fly data augmentation techniques were applied in both pipelines, according to a predefined probability. The augmentations and their respective probability of application were:

\begin{itemize}
 \item input channel rescaling: multiplying each voxel by a factor uniformly sampled between 0.9 and 1.1 (A: 80\% probability, B: 20\%).
 \item input channel intensity shift: Adding each voxel a constant uniformly sampled between -0.1 and 0.1 (A: not performed, B: 20\% probability).
 \item additive gaussian noise, using a centered normal distribution with a standard deviation of 0.1.
 \item input channel dropping: all voxel values of one of the input channels were randomly set to zero (A: 16\% probability, B: not performed).
 \item random flip along each spatial axis (A: 80\% probability, B: 50\%).
\end{itemize}

\subsection{Training details}

Models were produced by a five-fold cross-validation. The validation set was only used to monitor the network performance during training, and to benchmark its performance at the end of the training procedure.

\subsubsection{Pipeline A:} For each fold, the neural network was trained for 200 epochs with an initial learning rate of 1e-4, progressively reduced by a cosine decay after 100 epochs \cite{he_bag_2018}. A batch size of 1 and the Ranger optimizer \cite{liu_variance_2020,zhang_lookahead_2019,yong_gradient_2020} were used. After 200 epochs, we performed a training scheme inspired from the fast stochastic weight averaging procedure \cite{athiwaratkun_there_2019}. The initial learning rate was restored to half of its initial value (5e-5), and training was done for another 30 epochs with cosine decay. Every 3 epochs, the model weights were saved. This procedure was repeated 5 times for a total of 150 additional epochs. At the end, the saved weights were averaged, effectively creating a new “self-ensembled” model. The Adam optimizer \cite{kingma_adam_2017} was used without weight decay for the stochastic weight averaging procedure.
\subsubsection{Pipeline B:} The maximum number of training iterations was set to 400. The best model kept was the one with the lowest loss value on the validation set. A batch size of 3 and Adam optimizer with an initial learning of 1e-4 and no weight decay. Cosine annealing scheduler was used.
\subsubsection{Common:} In order to train a bigger neural network, float 16 precision (FP16) was used, which reduced memory consumption, accelerated the training procedure, and may lead to extra performance \cite{he_bag_2018}.

The neural network was built and trained using Pytorch v1.6 (which has native FP16 training capability) on Python 3.7. The model could fit on one graphic card (GPU).

\subsection{Inference}

Inference was performed in a two-steps fashion. First, models available from each pipeline were ensembled separately, by simple predictions averaging. Consequently, two labelmaps per case, one for each pipeline, were created. Three different models per fold (except one fold due to time constraint) were available for pipeline A: a 3D attention U-net version, a U-net version trained on an unfiltered version of the training dataset, and a U-net version trained on a filtered subset of the training dataset. The filtering process was based on previous training runs: cases with high training loss at the end of the training procedure were flagged as potentially wrong and removed from the complete training set, thus creating a “cleaned” version of the training dataset. The top two performing models per fold were chosen for ensembling (A). For Pipeline B, the five cross-validated models (one per fold) were ensembled. Then, the two labelmaps are merged based on the individual performance of each ensemble on the online validation set, as described below.

\subsubsection{First step}
For each pipeline, the initial volume was preprocessed like the training data, then cropped to the minimal brain extent, and finally zero-padded to have each of the spatial dimensions divisible by 8. Test time augmentation (TTA) was done using 16 different augmentations for each of the models generated by the cross-validation, for a total of 80 predictions per sample. We used flips, and 90-180-270 rotations only in the axial plane, as rotation in other planes led to worse performance on the local validation set. Final prediction was made by averaging the predictions, using a threshold of 0.5 to binarize the prediction.
Labelmap reconstruction was then performed in a straightforward manner: ET prediction was left untouched, the NET-NEC region of the tumor was deduced from a boolean operation between the ET label and the TC label, and similarly for the edema between the TC and the WT label ($NonEnhanching = TC – ET$ ; $edema = WT – TC$).

\subsubsection{Second step}

The first step gave two labelmaps per case. Based on the online validation dataset, the mean whole tumour dice metric of the pipeline B's ensemble was consistently higher than that of the pipeline A's ensemble. We hypothesized that models from pipeline B were better for predicting edema. To keep the score intact on ET and TC from models A, ET and NET/NCR predicted labels had to be left untouched. If A predicted background or edema and B predicted edema or background respectively, B predicted labels were kept. The merging procedure is shown in table \ref{merging}

\begin{table}
\caption{Merging procedure of the  two labelmaps. 0: background, 1: necrotic and non-enhancing tumor core (NET), 2: peri-tumoral edema (ED), 4:enhancing tumor (ET)}
    \centering
    \label{merging}
    \begin{tabular}{l|l|l|l|l|l|}
\cline{3-6}
\multicolumn{2}{l|}{\multirow{2}{*}{}}                               & \multicolumn{4}{l|}{Model A}                                                          \\ \cline{3-6} 
\multicolumn{2}{l|}{}                                                & \textit{\textbf{0}} & \textit{\textbf{1}} & \textit{\textbf{2}} & \textit{\textbf{4}} \\ \hline
\multicolumn{1}{|l|}{\multirow{4}{*}{model B}} & \textit{\textbf{0}} & 0                   & 1                   & 0                   & 4                   \\ \cline{2-6} 
\multicolumn{1}{|l|}{}                         & \textit{\textbf{1}} & 0                   & 1                   & \textbf{2}          & 4                   \\ \cline{2-6} 
\multicolumn{1}{|l|}{}                         & \textit{\textbf{2}} & \textbf{2}          & 1                   & 2                   & 4                   \\ \cline{2-6} 
\multicolumn{1}{|l|}{}                         & \textit{\textbf{4}} & 0                   & 1                   & 2                   & 4                   \\ \hline
\end{tabular}
\end{table}

\subsection{Ablation Study for Pipeline A}

Experiments with and without dataset filtering and attention block were produced for pipeline A. Cross-validated results can be found in Table \ref{tab3}. There was no clear benefit of either strategy, hence we decided to keep the two best available models for each fold for this pipeline.

\begin{table}
    \caption{Ablation study: results from cross-validation on the training set.}
    \label{tab3}
    \centering
    \begin{tabular}{|l|l|l|l|}
    \hline
    \multicolumn{1}{|c|}{Dice: mean(std)} & \multicolumn{1}{c|}{ET} & \multicolumn{1}{c|}{WT} & \multicolumn{1}{c|}{TC} \\
    \hline
    U-Net like & 0.8077 (0.011) & 0.9070 (0.006) & 0.8705 (0.013) \\
    + Patients removal & 0.8126 (0.019) & 0.9043 (0.005) & 0.8686 (0.012) \\
    + Attention block  & 0.8144 (0.022) & 0.9037 (0.008) & 0.8701 (0.018)  \\
    \hline
    \end{tabular}
\end{table}

\section{Results}
\subsection{Online Validation dataset}

\begin{figure}
    \centering
    \includegraphics[width=\textwidth]{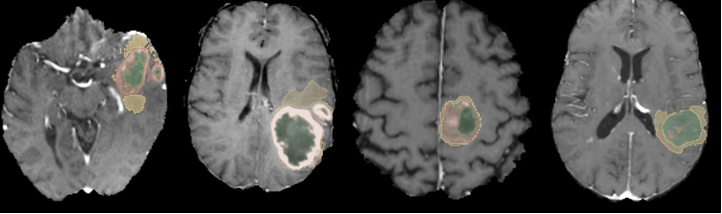}
    \caption{From left to right: ground truth example from the training set, and generated segmentations from our solution for three patients among the online validation set; respectively: best mean dice score (ET:0.95, WT:0.96, TC:0.98), average mean dice score (ET:0.73, WT:0.92, TC:0.93), and worst mean dice score (ET:0.23, WT:0.95, TC:0.13). \textbf{Red}: enhancing tumor (ET), \textbf{Green}: non enhancing tumor/ necrotic tumor (NET/NCR), \textbf{Yellow}: peritumoral edema (ED) } 
    \label{fig3}
\end{figure}

Table \ref{tab1} displays the results for the online validation data. Our models produced a Dice metric greater than 0.8. for each tumor region. Our two-pass merging strategy had no impact on the ET and TC segmentation performance of the pipeline A's ensemble, while greatly improving WT segmentation. Single pass strategy already yielded good performance for all three tumor regions. Larger value of Hausdorff distance for ET compared to other tumor subregions is explained by the absence of the ET label for some cases. Consequently, predicting even one voxel of ET would lead to a major penalty for this metric. 
Example of segmented tumor from the online validation set is displayed in Figure \ref{fig3}. It is hard to visually discriminate best from the average result, based on the mean dice score per patient (average across the three tumor sub-regions). However, our worst generated mask showed obvious error: contrast enhanced arteries were mislabeled as enhancing tumor.

\begin{table}
    \caption{Performance on the complete BraTs’20 Online Validation Data for the merging strategy, unless otherwise specified.}
    \centering
    \label{tab1}
    \begin{tabular}{|l|l|l|l|}
    \hline
    Metric (mean)              & ET       & WT      & TC      \\
    \hline
    Dice (Pipeline A alone)    & 0.80585  & 0.89518 & 0.85415 \\
    Dice (Pipeline B alone)    & 0.72738  & 0.91123 & 0.84921 \\
    \hline
    Dice                      & 0.80585  & 0.91148 & 0.85416 \\
    Sensitivity               & 0.81488  & 0.91938 & 0.84485 \\
    Specificity               & 0.99970  & 0.99915 & 0.99963 \\
    Hausdorff (95\%)          & 20.55756 & 4.30103 & 5.69298 \\
    \hline
    \end{tabular}
\end{table}

\subsection{Testing dataset}

Our final results on the testing dataset are displayed in table 2. These results ranked us among the top 10 teams for the segmentation challenge. A significant discrepancy between validation and testing datasets for the TC Hausdorff distance was visible, while all other metrics showed small but limited overfit.

\begin{table}
    \caption{Performance on the BraTs’20 Testing Data.}
    \label{tab2}
    \centering
    \begin{tabular}{|l|l|l|l|}
    \hline
    Metric (mean)    & ET       & WT      & TC       \\
    \hline
    Dice             & 0.78507  & 0.88595 & 0.84273  \\
    Sensitivity      & 0.81308  & 0.91690 & 0.85934  \\
    Specificity      & 0.99967  & 0.99905 & 0.99964  \\
    Hausdorff (95\%) & 20.36071 & 6.66665 & 19.54915 \\
    \hline
    \end{tabular}
\end{table}

\section{Discussion}

Our solution to the BraTS'20 challenge is based on standard approaches carefully crafted together: we used U-net 3D neural networks, trained with on-the-fly data augmentations using the Dice Loss and deep supervision, and inferred using test time augmentation and models predictions ensembling.

Many modern “bells and whistles” were tried: short additive residual connections \cite{he_deep_2015}, dense blocks \cite{huang_densely_2018}, more recent neural networks backbone based on inverted residual bottleneck \cite{howard_mobilenets_2017}, newer decoder structure like biFPN layer \cite{tan_efficientdet_2020}, or semi-supervised setting using brain dataset from the Medical Decathlon \cite{simpson_large_2019}. None of these refinements led to significant improvement on the local validation set. We hypothesize that this was probably due to GPU memory constraints. Indeed, while these layers improve the model accuracy at a relatively small parameter cost, it increases significantly the size of the activation maps of the model, forcing us to use smaller networks (reduction of the number of output channels per convolutional layer). Reducing the crop size of the patch was not an option as this would have most probably reduced the network performance due to the lack of context. Moreover, all of these additions led to a significant increase of the training time, reducing the searchable space in the limited timeframe of the challenge.

Stochastic weight averaging at the end of the training was the most notable refinement we used. This training scheme was a remnant from the mean teacher semi-supervised training \cite{tarvainen_mean_2018}. We did not benchmark its real potential but expect it to produce a more generalizable model, to prevent from overfitting on the training set and to remember the noisy labels. Indeed, it has been shown that a high learning rate could prevent such behavior, and we expect that our training benefits from the multiple learning rate restarts \cite{tanaka_joint_2018}.

\begin{figure}
    \centering
    \includegraphics[width=8cm]{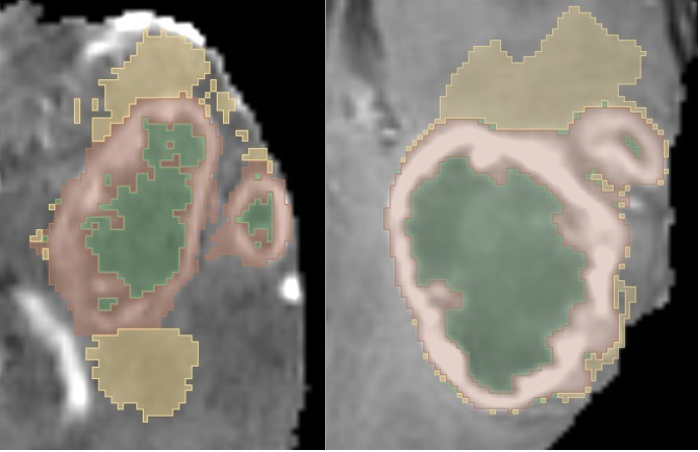}
    \caption{Zoomed version of the first two vignettes of Figure \ref{fig3} Left: ground truth example from the training set. Right: generated segmentations from our solution for the best mean dice score patient on the validation set. \textbf{Red}: enhancing tumor (ET), \textbf{Green}: non enhancing tumor/ necrotic tumor (NET/NCR), \textbf{Yellow}: peritumoral edema (ED). It is interesting to note that both exhibit the same pattern: central non enhancing tumor core with surrounding enhancing ring and diffuse peritumoral edema.} 
    \label{zoom_in}
\end{figure}

Notably, while our results were not state of the art for the BraTS 2020 challenge, the segmentation performance of our method is in the usual range of inter-rater agreement for lesion segmentation \cite{chassagnon_ai-driven_2020,tacher_semiautomatic_2013} and could already be valuable for clinical use. As an example, Figure \ref{zoom_in} zooms in the tumor segmentation of the first two annotations of Figure \ref{fig3} (respectively manual ground truth annotations and best validation case).

\section{Conclusion}

The task of brain tumor segmentation, while challenging, can be solved with good accuracy using 3D U-Net like neural network architecture, with a carefully crafted pre-processing, training and inference procedure. We open-sourced our training pipeline at \url{https://github.com/lescientifik/open_brats2020}, allowing future researchers to build upon our findings, and improve our segmentation performance.

\clearpage
\bibliographystyle{splncs04}
\bibliography{ref_bibtex}

\end{document}